\documentclass[11pt]{article}

\usepackage{acl}

\usepackage{times}
\usepackage{latexsym}

\usepackage[T1]{fontenc}

\usepackage[utf8]{inputenc}

\usepackage{microtype}

\usepackage{inconsolata}

\usepackage{graphicx}

\usepackage{xspace}
\usepackage{dsfont}
\usepackage{amsmath}
\usepackage{booktabs}
\usepackage{subcaption}
\usepackage{amsfonts}
\usepackage{pifont}
\usepackage{xcolor}
\usepackage[most]{tcolorbox}
\usepackage{listings}
\tcbuselibrary{breakable}

\newcommand{\modelname}{MUSE\xspace}

\title{\modelname: Revealing Responsiveness Gaps of LLM Tutors on Multiple-Solution Programming Problems}

\author{
  Sky Cen\\
  University of Central Florida\\
  \texttt{sky.cen@ucf.edu} \\
  \And
  Shashank Sonkar\\
  University of Central Florida \\
  \texttt{shashank.sonkar@ucf.edu} \\
  \\
}

\begin{document}

\maketitle

\begin{abstract}
When a problem admits multiple valid solutions, effective tutoring requires building on a student's strategy while introducing alternatives at appropriate points. 
However, whether LLM tutors can support this flexibility by appropriately responding to students’ diverse solution approaches remains largely unexplored.
In this work, we introduce \modelname: Multi-solution Understanding and Simulation Environment, a controlled environment for simulating student-tutor conversations on multiple-solution programming problems.
Grounded in learning science on responsive instruction and multiple-solution learning, we investigate which solutions LLM tutors prefer, how they respond to students' approaches, and how they transition between solutions. 
Experiments suggest that LLM tutors exhibit systematic preferences among valid solutions, often favoring brute-force approaches when not providing general advice. 
More concerningly, tutors often impose these preferences, redirecting students toward their own preferred approach rather than supporting the student's.
Yet this tendency reverses once a solution is completed: tutors often remain anchored to the existing approach when a stronger alternative could deepen understanding, and sometimes even transition to a less optimal approach.
Thus, tutors struggle both to build on students' reasoning when appropriate and to extend their reasoning when beneficial.
Promisingly, targeted post-training improves both tutor responsiveness and transitions toward more optimal solutions.
\end{abstract}

\section{Introduction}

\begin{figure*}[!t]
    \centering
    \includegraphics[width=\linewidth]{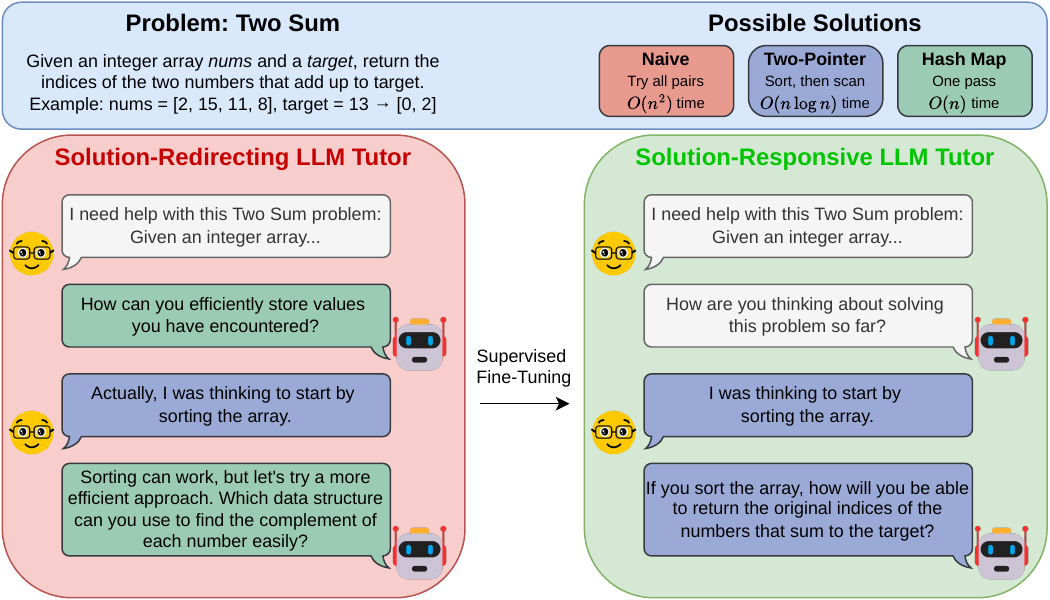}
    \caption{Comparison between a solution-redirecting LLM tutor (left) and a solution-responsive LLM tutor (right). The solution-redirecting tutor acknowledges the student's valid but suboptimal two-pointer approach, yet redirects them towards the more optimal hash map solution rather than supporting their reasoning. In contrast, the solution-responsive tutor builds on the student's approach through a pedagogically guided question.}
    \label{fig:teaser}
\end{figure*}

Large language models (LLMs) are increasingly used as interactive tutors, leveraging their ability to generate explanations, ask questions, provide feedback, and engage learners in dialogue. Recent work has explored LLM-based tutoring across dimensions including pedagogical alignment, misconception awareness, and metacognitive support \cite{macina-etal-2023-mathdial, Scarlatos_2025, wang2026mirror}. However, existing LLM tutoring systems and evaluations largely treat the target solution as fixed, or focus on a particular solution even when a problem admits multiple valid approaches. They therefore do not explicitly model the student’s chosen solution strategy as a central part of the tutoring interaction.

This distinction is especially important for programming problems, where a single problem can admit multiple correct solutions with different tradeoffs in efficiency, simplicity, and generalizability. A \textbf{responsive} tutor must do more than guide a student toward a correct answer: it must recognize the approach the student is pursuing, determine how to build on that reasoning, and decide when and how to transition to an alternative strategy \cite{_atman_Aksoy_2026, Dyer_2015}. This raises a central question: \textbf{Can LLM tutors effectively support students across multiple-solution strategies?}

In this work, we introduce \modelname (Multi-solution Understanding and Simulation Environment), a controlled environment for simulating tutoring interactions on multiple-solution programming problems. \modelname explicitly controls the solution pursued by the student while tracking the solution intended by the tutor at each time step, enabling systematic evaluation of how LLM tutors respond to different student approaches. Using this environment, we investigate three research questions: (RQ1) which solution strategies LLM tutors prefer when multiple approaches are available, (RQ2) whether tutors respond to and build upon the student’s proposed solution, and (RQ3) whether tutors appropriately transition between solution strategies after the student completes an approach.

Comprehensive experiments across LLM tutors spanning GPT-OSS, Qwen, Llama, and Gemma families reveal that tutors’ opening responses tend to provide general advice rather than guidance toward a specific solution. When tutors provide solution-specific guidance, they exhibit systematic preferences, frequently favoring brute-force or otherwise naive approaches. 

Strikingly, tutors do not simply exhibit these preferences, but often impose them. Even when students pursue valid alternative strategies, tutors frequently redirect them towards the tutor's preferred approach instead (Figure~\ref{fig:teaser}). Yet this tendency reverses after the student has completed their solution: when given the opportunity to deepen the student's understanding with a stronger alternative solution, tutors often remain anchored to the existing approach. Even worse, tutors may guide students towards a less optimal approach. 

Together, these results reveal a broader failure of solution responsiveness: LLM tutors struggle to both support a student's reasoning when they should and to move beyond it when appropriate. Finally, we show that targeted supervised fine-tuning on interactions that explicitly demonstrate responsive behavior improves both tutor responsiveness and transitions toward more optimal solutions, suggesting that these tutoring behaviors can be learned through post-training.

Our main contributions are threefold:
\begin{enumerate}
    \item We introduce \modelname together with evaluation metrics for measuring solution preferences, responsiveness, and transitions between solution strategies.
    
    \item We systematically evaluate open-source LLM tutors on multiple-solution programming problems, revealing limitations in their responsiveness to students' proposed solutions and their transitions between solution strategies.
    
    \item We demonstrate that targeted post-training can mitigate these limitations, leveraging \modelname to generate synthetic data at scale.

\end{enumerate}

\section{Related Work}

\subsection{LLMs for Tutoring}

Prior work has investigated LLM tutoring through pedagogical alignment \cite{dinucu-jianu-etal-2025-problem, Scarlatos_2025}, misconception awareness \cite{macina-etal-2023-mathdial, Otero_2025}, metacognitive development \cite{liu2026metaclass, wang2026mirror}, and personalized instructional support \cite{Reddig_2025}. Despite these advances, existing evaluations commonly focus on whether a tutor provides correct or pedagogically appropriate guidance toward a target solution, even when the underlying problem admits multiple valid approaches \cite{macina-etal-2023-mathdial, Reddig_2025, Scarlatos_2025}. 
As a result, relatively little attention has been given to whether a tutor recognizes and builds upon the particular solution pursued by a student, or how it should transition between alternative solution strategies. Our \modelname allows for evaluating tutors’ solution preferences, responsiveness, and transitions between solution strategies by explicitly controlling the student’s proposed solution.

\subsection{Multiple-Solution Problems}

Educational research shows that engaging with multiple-solution strategies can promote learning, conceptual understanding, and procedural flexibility \cite{GROE2006122, Rittle_Johnson_2007, Rittle_Johnson_2009}. Because multiple approaches can provide distinct learning opportunities, understanding how tutors navigate this solution space is an important first step. We therefore examine which valid solutions LLM tutors prefer (\textbf{RQ1}).

However, simply presenting multiple approaches may overwhelm students \cite{Lynch_2014}, making it important for instructors to recognize and build on students’ existing reasoning \cite{_atman_Aksoy_2026, Choppin_2011, Dyer_2015}. This motivates \textbf{RQ2}: whether LLM tutors support students’ proposed solutions rather than redirecting them toward another approach.

Multiple-solution instruction also creates opportunities to introduce alternatives after students complete an initial solution. Because explanations are most effective when learners have reached an appropriate point for receiving them \cite{Schwartz_1998}, we investigate whether LLM tutors capitalize on these opportunities by transitioning toward a more optimal solution, continuing with the existing approach, or redirecting students toward a less optimal solution (\textbf{RQ3}).

These findings suggest that effective tutoring across multiple solutions requires navigating students’ reasoning and solution strategies throughout the interaction. However, LLM tutoring research has not systematically studied these behaviors. We therefore study whether LLM tutors can effectively support students across multiple-solution strategies. A detailed discussion of related work on multiple-solution instruction is provided in Appendix~\ref{appendix:related work}.

\section{\modelname: Multi-solution Understanding and Simulation Environment}
\begin{figure*}[t]
    \centering
    \includegraphics[width=\linewidth]{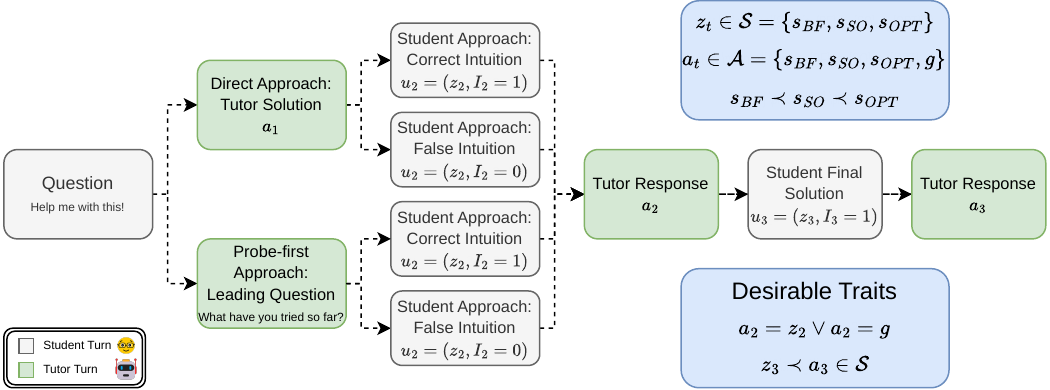}
    \caption{Conversation trajectories supported by \modelname. \modelname simulates student-tutor conversations with controlled tutor openings and student solution intuitions. We use \modelname to systematically measure tutor solution preferences, responsiveness, and transitions between solution approaches. \modelname is adapted to generate synthetic solution-responsive conversations that explicitly guide students toward more optimal solutions for targeted post-training.}
    \label{fig:pipeline}
\end{figure*}

\modelname provides a controlled environment for simulating LLM tutoring interactions on programming problems with multiple-solution approaches. It controls the simulated student’s pursued solution and intuition, as well as the tutor’s conversation initialization, enabling systematic evaluation of tutor behavior across different learning trajectories at scale. Using these controls, we define a response taxonomy and evaluation metrics for tutor's solution preference, responsiveness, and transitions between solution strategies (Figure~\ref{fig:pipeline}). 

\subsection{Task Formulation}
Let a programming problem be $p$, with three solution approaches: a brute force, a sub-optimal, and an optimal solution: $\mathcal{S} = \{s_{BF}, s_{SO}, s_{OPT}\}$. We define the student's turn as $u_t = (z_t, I_t)$, where $z_t \in \mathcal{S}$ is the solution approach pursued by the student and $I_t \in \{1, 0\}$ denoting whether the student's current intuition is correct, at each time step $t$. The conversation is then defined as follows:
\begin{equation}
    C_t = (p, u_1, a_1,\ldots, u_t, a_t),
\end{equation}
where the LLM tutor response is defined as 
\begin{equation}
    a_t \sim \pi_\theta(\cdot \ | \ C_t^-),
\end{equation}
where $\pi_{\theta}$ is the LLM tutor and 
\begin{equation}
    C_t^- = (p, u_1, a_1,\ldots, u_t).
\end{equation}

\subsection{Tutor Response Taxonomy}
\label{sec:tutor response taxonomy}
At all time steps $t > 0$, an LLM judge (Appendix~\ref{appendix:judge}) classifies the tutor's response into four main categories based on the tutor's prescribed approach:
$a_t \in \mathcal{A} = \{s_{BF}, s_{SO}, s_{OPT}, g\}$, where $g$ represents general advice that applies to the problem regardless of the student's solution.

\subsection{Controlled Student Simulation}
For each problem, \modelname allows individual student solutions to be simulated. To measure responsiveness, \modelname ensures that the student's proposed solution differs from the tutor's initial preferred solution (i.e. $z_2 \neq a_1$).

Additionally, the student's intuition can be independently controlled by the simulator: $I_t \in \{1, 0\}$ where $1$ denotes a correct intuition and $0$ indicates an incorrect intuition on a specific solution. This enables simulation of conversations where a student pursues a valid solution while having a flawed justification.

\subsection{Conversation Initialization}
\label{sec:conv init}
When simulating conversations, the LLM tutor will often open up the conversation by guiding the student towards a specific solution or by offering general advice. Alternatively, \modelname also allows for the LLM tutor to take a probe-first approach and open with a leading question \cite{cole1978mind}. More specifically, after the student asks for help with a problem, the tutor may respond with a question such as: ``What have you tried so far?" 

\subsection{Evaluation Framework}
\modelname enables systematic evaluation of three research questions: (RQ1) tutor solution preferences, (RQ2) responsiveness to students’ proposed solutions, and (RQ3) transitions between solution strategies.

\subsubsection{RQ1: LLM Tutor Solution Preference}
\label{sec:rq1}
The tutor’s initial instructional strategy reveals whether it exhibits a systematic preference for particular solution approaches before observing the student’s pursued solution. Furthermore, it provides a baseline for distinguishing the tutor’s intrinsic solution preferences from its subsequent responsiveness to the student’s approach. To measure the LLM tutor's solution preference, \modelname analyzes the tutor's opening response ($t = 1$), $a_1$. Across $N$ conversations, the empirical distribution over the tutor's opening response categories is estimated as:
\begin{equation}
    P(a_1 = a) = \frac{1}{N} \sum_{i=1}^N \mathds{1}[a_1^i = a], \ a \in \mathcal{A}.
\end{equation}
This gives the proportion of conversations in which the tutor's opening response targets a brute force, suboptimal, or optimal solution or provides general advice.

\subsubsection{RQ2: LLM Tutor Responsiveness}
\label{sec:rq2}
In instructional settings, tutors are expected to have a certain level of responsiveness towards their students. In multiple-solution problems, this implies that tutors should appropriately address and support a student's approach, regardless of their intuition \cite{Jacobs_2015}. We define the responsiveness of the tutor at turn $t > 1$ as:
\begin{equation}
    r_t = \mathds{1}[a_t = z_t \space \vee \space a_t = g] \in \{0, 1\}.
\end{equation}
More specifically, the tutor is responsive at a specific turn if the solution they are guiding the student towards is the same as the student's intended approach. Additionally, the tutor is also considered responsive if they give the student general advice, as general advice applies to every solution.

For RQ2, we measure the Solution Support Rate (SSR) of the tutor at turn $t = 2$ across $N$ conversations:
\begin{equation}
    \label{eq:ssr}
    \mathrm{SSR}_2 = \frac{1}{N} \sum_{i = 1}^N r_2^i = \frac{1}{N} \sum_{i = 1}^N \mathds{1}[a_2^i = z_2^i \space \vee \space a_2^i = g].
\end{equation}
In other words, this is the percentage of conversations that tutor supports or provides general advice the turn after the student offers their approach. 

\subsubsection{RQ3: Post-Solution Transitions}
\label{sec:rq3}
After a student has solved the problem using their approach, the tutor is presented with an opportunity to teach the student a more optimal approach. In the context of programming problems, such approaches can involve distinct tradeoffs in efficiency, simplicity, or generalizability. We define a ranking system as follows: 
\begin{equation}
    \begin{cases}
        \operatorname{rank}(s_{BF}) = 1 \\
        \operatorname{rank}(s_{SO}) = 2 \\
        \operatorname{rank}(s_{OPT}) = 3
    \end{cases}
\end{equation}
The ranking system creates a solution hierarchy based on solution efficiency. More specifically: 
\begin{equation}
    \operatorname{rank}(s_{BF}) < \operatorname{rank}(s_{SO}) < \operatorname{rank}(s_{OPT}),
\end{equation}
or equivalently:
\begin{equation}
    s_{BF} \prec s_{SO} \prec s_{OPT}.
\end{equation}
Using the ranking system, we quantify the direction of a tutor's transition at time step $t > 1$ as:
\begin{equation}
    \Delta_t =
    \begin{cases}
        \operatorname{sgn}\!\left(
            \operatorname{rank}(a_t) - \operatorname{rank}(z_t)
        \right), & \text{if } a_t \in \mathcal{S} \\[4pt]
        0, & \text{if } a_t = g.
    \end{cases}
\end{equation}
where $\operatorname{sgn}(\cdot)$ denotes the sign function, mapping positive, zero, and negative values to $1, 0, -1$ respectively. Thus, $\Delta_t = 1$ indicates that the tutor transitions to a more optimal solution, $\Delta_t = -1$ indicates a transition to a less optimal solution, and $\Delta_t = 0$ indicates that the tutor remains at the same solution approach.

For RQ3, we evaluate the tutor's tendency to transition to more optimal solutions after the student has completed their solution at $t = 3$ across $N$ conversations:
\begin{equation}
    \label{eq:pts}
    \begin{aligned}
        \mathrm{PTS}_3 &= \frac{1}{N} \sum_{i = 1}^N \Delta_3^i
    \end{aligned}
\end{equation}
We refer to this metric as the Post-Solution Transition Score (PTS). This metric is bounded between $[-1, 1]$, attaining its maximum when the tutor transitions to a more optimal solution in every conversation and vice versa. A value of $0$ indicates that, on average, the tutor makes no net change in solution optimality after the student completes their solution.

\section{Post-Training Methodology}
In this section, we propose a methodology for improving LLM tutor responsiveness through SFT by generating a dataset with examples of explicit tutor responsiveness and transitions towards more optimal solutions.

\subsection{Synthetic Dataset Generation}
\label{sec:synthetic dataset generation}

We adapt \modelname to generate conversations that explicitly demonstrate responsive behavior across all possible solution trajectories. Rather than sampling tutor responses from the LLM tutor $\pi_\theta$, we use a separate response generator $\pi_{\textrm{gen}}$. The generator is provided with the problem $p$, the solution set $\mathcal{S}$, and the controlled learner trajectory, and is prompted to produce tutor responses satisfying our desired behavioral constraints.

For a given problem $p$, the learner trajectory is defined as
\begin{equation}
    U = (u_1, \ldots, u_T), \ u_t = (z_t, I_t),
\end{equation}
and generated tutor responses are denoted 
\begin{equation}
    \hat{a}_t \sim \pi_{\textrm{gen}}(\cdot \ | \ \hat{C}_t^-, \mathcal{S})
\end{equation}
where 
\begin{equation}
    \hat{C}_t^- = (p, u_1, \hat{a}_1, \ldots, u_t).
\end{equation}
Thus, the synthetic conversation is defined as:
\begin{equation}
    \hat{C}_t = (p, u_1, \hat{a}_1, \ldots, u_t, \hat{a}_t).
\end{equation}

\modelname initializes the generated conversations using the probe-first approach from \S~\ref{sec:conv init}. This is because by opening with a guiding question rather than jumping right into a solution, the tutor is better able to judge the current understanding of the student \cite{cole1978mind, Michaels_2007}. This is particularly important for multiple-solution problems where certain solutions may be too challenging for the student based on their current knowledge \cite{Schwartz_1998}.

To encourage solution-responsive behavior, we constrain the generated responses at each stage of the trajectory. After the student specifies their approach, the generator produces a pedagogical response pertaining to the student's solution:
\begin{equation}
    \hat{a}_2 = z_2
\end{equation}
After the student completes their solution, the generator produces a response that acknowledges the student's approach and transitions towards a more optimal solution:
\begin{equation}
    z_3 \prec \hat{a}_3 \in \mathcal{S}    
\end{equation}
The resulting synthetic conversations provide demonstrations of the desired tutor behavior in a synthetic dataset:
\begin{equation}
    \mathcal{D}_\mathrm{SFT} = \{(\hat{C}_t^-, \hat{a}_t)\}_{t = 1}^{T}.
\end{equation}
The complete generation prompts are provided in Appendix~\ref{appendix:responsive prompt} and \ref{appendix:optimal transition prompt}.

\subsection{Supervised Fine-Tuning}
\label{sec:sft}
We perform supervised fine-tuning (SFT) to train LLM tutors to exhibit solution-responsive behavior. Using the synthetic dataset $\mathcal{D}_\mathrm{SFT}$, SFT optimizes the standard autoregressive language modeling objective:
\begin{equation}
\mathcal{L}_{\mathrm{SFT}}(\theta) = -\sum_{(\hat{C}_t^-, \hat{a}_t) \in \mathcal{D}_\mathrm{SFT}} \log \pi_\theta(\hat{a}_t \ | \ \hat{C}_t^-).
\end{equation}
The training examples explicitly demonstrate solution-responsive tutoring, allowing the model to learn how to adapt its responses to the student’s reasoning.

We additionally consider a solution-informed SFT setting, in which the candidate solution space $\mathcal{S}$ is provided to the model as context through the system prompt. The corresponding objective is:
\begin{equation}
    \label{eq:sft solutions}
    \mathcal{L}_{\mathrm{SFT}}(\theta) = -\sum_{(\hat{C}_t^-, \hat{a}_t) \in \mathcal{D}_\mathrm{SFT}} \log \pi_\theta(\hat{a}_t \ | \ \hat{C}_t^-, \mathcal{S}).
\end{equation}
This setting provides the model with explicit information about the available solution approaches in addition to the tutoring context.

For post-training, we generate the synthetic training data described in \S~\ref{sec:synthetic dataset generation} using only the training split. We fine-tune Gemma 4 12B IT model for each SFT loss function. We consider three SFT variants that differ in whether candidate solutions are provided as system-prompt context during training and inference (Table~\ref{tab:sft models}). \emph{Gemma SFT1} is trained and evaluated without candidate solutions. \emph{Gemma SFT2} is trained with candidate solutions but evaluated without them, while \emph{Gemma SFT3} receives candidate solutions during both training and inference. This setup allows us to isolate the effect of providing candidate solutions as contextual information to the tutor.

\begin{table}[t]
    \centering
    \footnotesize
    \setlength{\tabcolsep}{4pt}
    \begin{tabular}{lcc}
        \toprule
        Model & Training $\mathcal{S}$ & Inference $\mathcal{S}$ \\
        \midrule
        Gemma & \ding{53} & \ding{53} \\
        Gemma SFT1 & \ding{53} & \ding{53} \\
        Gemma SFT2 & \ding{51} & \ding{53} \\
        Gemma SFT3 & \ding{51} & \ding{51} \\
        \bottomrule
    \end{tabular}
    \caption{Overview of SFT models. The table shows which SFT models have access to candidate solutions in the system prompt during training and inference time.}
    \label{tab:sft models}
\end{table}

\section{Experiments and Results}

\begin{table*}[t]
    \centering
    \noindent
    \begin{minipage}[t]{0.48\textwidth}
        \centering
        \fontsize{10pt}{11.5pt}\selectfont
        \setlength{\tabcolsep}{3.5pt}

        \begin{tabular}{l @{\hspace{4pt}} cccc}
            \toprule
            & \multicolumn{2}{c}{Correct Intuition}
            & \multicolumn{2}{c}{False Intuition} \\
            Model & Direct & Probe-first & Direct & Probe-first \\
            \midrule
            GPT-OSS   & 0.336 & 0.534 & 0.347 & 0.423 \\
            Qwen      & \textbf{0.472} & 0.589 & \textbf{0.514} & 0.544 \\
            Llama     & 0.465 & 0.459 & 0.413 & 0.279 \\
            Gemma     & 0.380 & 0.507 & 0.418 & 0.419 \\
            Gemma SFT1 & - & \textbf{0.606} & - & 0.596 \\
            Gemma SFT2 & - & 0.518 & - & 0.614 \\
            Gemma SFT3 & - & 0.568 & - & \textbf{0.619} \\
            \bottomrule
        \end{tabular}

        \captionof{table}{Quantitative results for RQ2: LLM tutor responsiveness. Solution Support Rate (SSR $\uparrow$) measures how often tutors support students’ proposed solutions. SFT notably increases SSR.}
        \label{tab:rq2}
    \end{minipage}
    \hfill
    \begin{minipage}[t]{0.48\textwidth}
        \centering
        \fontsize{10pt}{11.5pt}\selectfont
        \setlength{\tabcolsep}{3.5pt}

        \begin{tabular}{l @{\hspace{4pt}} cccc}
            \toprule
            & \multicolumn{2}{c}{Correct Intuition}
            & \multicolumn{2}{c}{False Intuition} \\
            Model & Direct & Probe-first & Direct & Probe-first \\
            \midrule
            GPT-OSS   & 0.393 & 0.296 & 0.363 & 0.375 \\
            Qwen      & 0.172 & 0.092 & 0.130 & 0.132 \\
            Llama     & \textbf{0.581} & 0.412 & \textbf{0.405} & 0.357 \\
            Gemma     & 0.222 & 0.343 & 0.300 & 0.349 \\
            Gemma SFT1 & - & 0.570 & - & 0.625 \\
            Gemma SFT2 & - & 0.573 & - & 0.498 \\
            Gemma SFT3 & - & \textbf{0.936} & - & \textbf{0.903} \\
            \bottomrule
        \end{tabular}

        \captionof{table}{Quantitative results for RQ3: post-solution transitions. Post-Solution Transition Score (PTS $\uparrow$) measures transitions toward more optimal solutions. SFT with candidate solutions significantly improves PTS.}
        \label{tab:rq3}
    \end{minipage}
\end{table*}

\subsection{Dataset and Experimental Setup}
\label{sec:dataset and experimental setup}

We collect 150 programming problems from the NeetCode 150 collection \cite{neetcode150}, spanning diverse topics and three difficulty levels: easy, medium, and hard. For each problem, we collect its title, description, and three valid solutions. Each solution includes an explanation, algorithm steps, and time complexity, and is categorized as \emph{brute-force}, \emph{suboptimal}, and \emph{optimal} based on time complexity. For each solution, we use GPT-OSS 120B \cite{agarwal2025gpt} to generate a correct intuition and a plausible but incorrect intuition that a student pursuing the solution may have (Appendix~\ref{appendix:correct intuition prompt},~\ref{appendix:false intuition prompt}). 

Using \modelname, we simulate one conversation per problem and conversation type with four LLMs as tutors: GPT-OSS 120B, Qwen3.8-27B \cite{qwen3.8}, Llama 3.3 70B Instruct \cite{grattafiori2024llama}, and Gemma 4 12B IT \cite{gemmateam2026gemma4}, yielding 150 conversations per type. GPT-OSS 120B serves as the LLM judge, classifying tutor responses according to the response taxonomy described in \S~\ref{sec:tutor response taxonomy} (Appendix~\ref{appendix:judge}). For each LLM tutor, \modelname generates conversations under four conditions: probe-first or direct opening, crossed with a correct ($I_2=1$) or false ($I_2=0$) student intuition. When presenting an approach, the simulated student draws on the corresponding generated intuition.

For SFT experiments, we randomly split the problems into 80\% training and 20\% test sets, using the training split for synthetic dataset generation and post-training and reserving the held-out test split for evaluation. To assess the robustness of SFT results, we repeat the training and evaluation process across three random problem splits generated using different seeds and report the mean of each metric (Appendix~\ref{appendix:synthetic dataset},~\ref{appendix:sft details}).

\subsection{RQ1: LLM Tutor Solution Preference}

\begin{figure}[t]
    \centering
    \includegraphics[width=\linewidth]{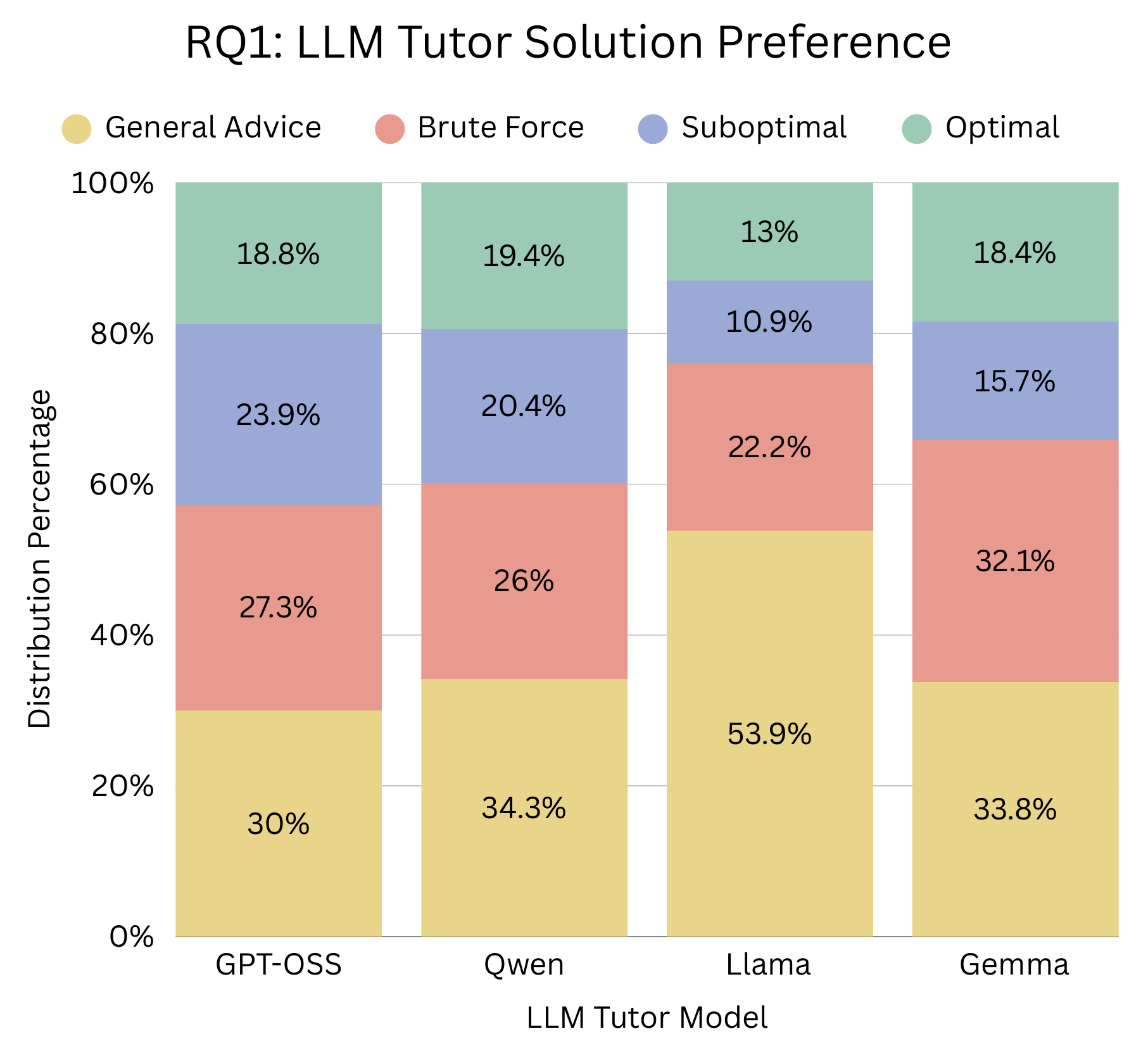}
    \caption{LLM tutor opening response distribution. LLM tutors favor brute-force approaches when not providing general advice.}
    \label{fig:rq1}
\end{figure}

We examine which solution approaches LLM tutors tend to favor when multiple correct approaches are available. We aggregate conversations with correct or false student intuitions where the tutor’s opening response is direct (solution-specific or general advice). This allows for a larger sample, yielding the empirical distribution described in \S~\ref{sec:rq1}.

We observe two notable patterns in LLM tutors' solution preferences, as revealed in Figure~\ref{fig:rq1}. First, among solution-specific opening responses, brute-force is consistently the most preferred approach across all four models ($22.2$ to $32.1\%$), while the most optimal approach accounts for only $13.0$ to $19.4\%$ of responses. Thus, LLM tutors do not consistently prioritize computationally optimal solutions, potentially because brute-force approaches offer a simpler starting point for explanation or are more prevalent in programming-oriented training data.

Second, models differ substantially in their propensity to provide solution-specific guidance: Llama provides general advice in $53.9\%$ of cases, compared with $30.0\%$ for GPT-OSS. Thus, model differences extend beyond solution preference to whether tutors commit to a particular solution at all. These findings provide a basis for evaluating tutor responsiveness: do tutors effectively adapt to student's approach when it differs from their preferred approach?

\subsection{RQ2: LLM Tutor Responsiveness}

We investigate whether LLM tutors genuinely support the solution proposed by the student. We report the Solution Support Rate (SSR) (Equation~\ref{eq:ssr}) for each conversation type across all four models. For fine-tuned models, we evaluate only the probe-first setting, since they are trained on this opener.

Table~\ref{tab:rq2} shows that probe-first openings generally increase SSR, with effects varying substantially across models. GPT-OSS, Qwen, and Gemma exhibit higher SSR under probe-first openings for correct student intuitions, with improvements of $19.8$, $11.7$, and $12.7$ percentage points, respectively. This is intuitive, as the tutor has not yet introduced its preferred solution. However, eliciting the student’s approach alone does not guarantee solution-responsive behavior, particularly when the student’s reasoning is incorrect.

\paragraph{Post-Training Improvements.} We find that targeted SFT improves solution responsiveness. Compared with base Gemma, Gemma SFT1, trained and evaluated without explicit solution information, increases SSR from $0.507$ to $0.606$ for correct intuitions and from $0.419$ to $0.596$ for false intuitions, corresponding to improvements of $9.9$ and $17.7$ percentage points, respectively. 

However, providing the candidate solution space during SFT does not consistently improve responsiveness across intuition types. SSR is highest for correct intuitions when candidate solutions are excluded from training, whereas including them during both training and inference yields higher SSR for false intuitions. This suggests that candidate solutions alone are insufficient to improve tutor responsiveness, as tutors must also learn to adapt their guidance based on the student's reasoning.

\subsection{RQ3: Post-Solution Transitions}
\label{sec:rq3 results}

\begin{figure}[t]
    \centering
    \includegraphics[width=\linewidth]{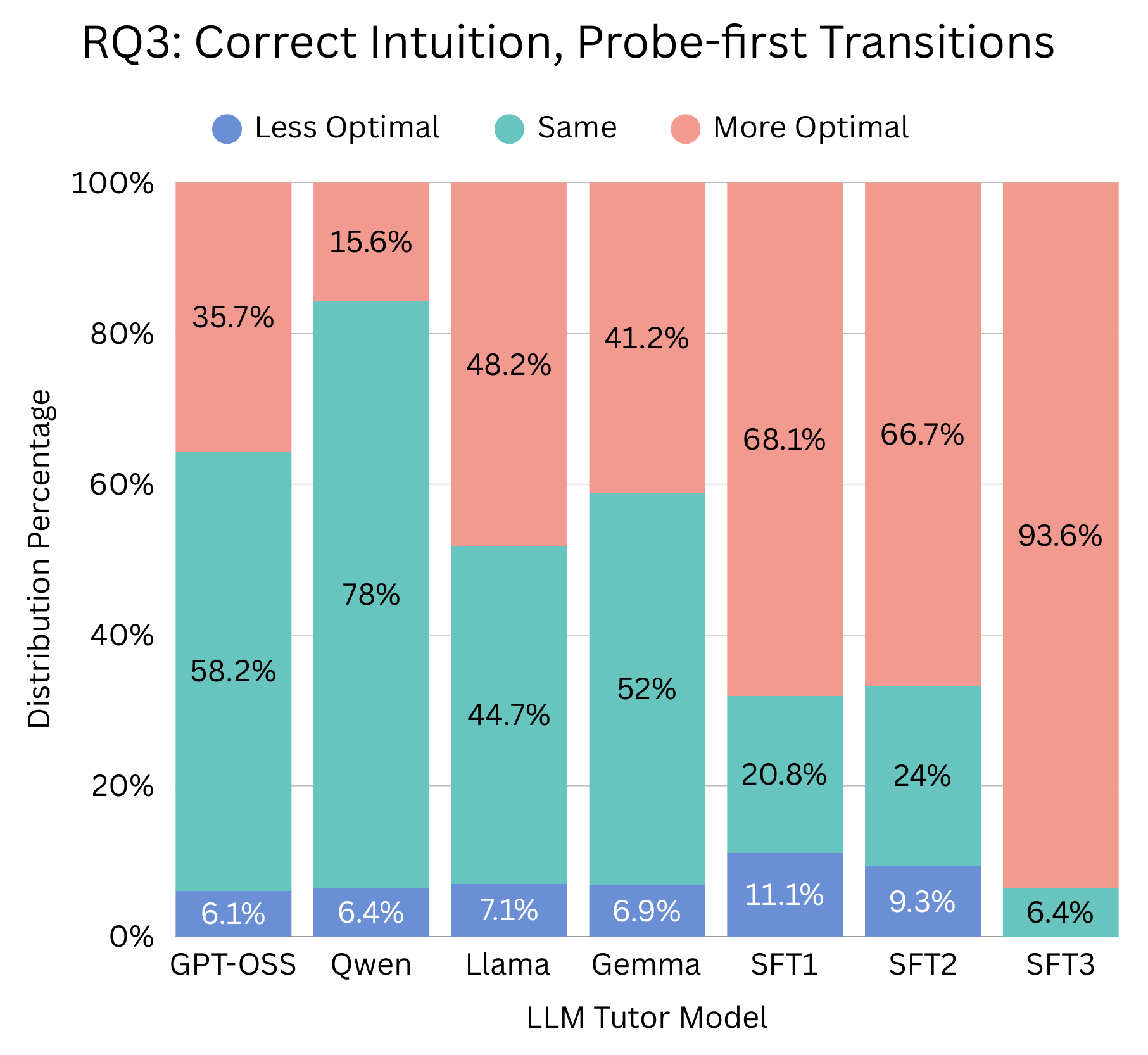}
    \caption{Correct intuition, probe-first transition distributions. Notably, all models demonstrate transitions towards less optimal solutions, however, SFT with candidate solutions in context during training and inference significantly mitigate these transition types.}
    \label{fig:rq3}
\end{figure}

We examine whether LLM tutors take advantage of the opportunity to introduce a more optimal solution after students complete their approach. We report the Post-solution Transition Score (PTS) (Equation~\ref{eq:pts}) for each conversation type across all four models. For fine-tuned models, we evaluate only the probe-first setting, since they are trained on this opener. Additionally, we present the transition distributions for correct intuition, probe-first approach (Figure~\ref{fig:rq3}) and other conversation types (Appendix Figures~\ref{fig:rq3_2},~\ref{fig:rq3_3},~\ref{fig:rq3_4}).

Table~\ref{tab:rq3} reveals substantial differences in LLM tutors’ tendencies to transition toward more optimal solutions. All base models exhibit positive PTS across conditions, indicating a general tendency toward more efficient solutions, but the magnitude varies considerably. Llama shows the strongest preference for post-solution optimization, while Qwen remains more anchored to the student’s solution. Despite Qwen’s relatively high SSR in RQ2, its low PTS in RQ3 shows that supporting a student’s approach does not necessarily lead to introducing a more optimal alternative after the student finishes their approach. More concerningly, tutors can steer students toward less optimal solutions around $6.1\%$ to $7.1\%$ of the time, potentially undermining their existing progress by encouraging less efficient approaches.

\paragraph{Post-Training Improvements.} The SFT results demonstrate that optimal solution transitions can be learned through explicit examples in the dataset. Accordingly, Gemma SFT1 increases PTS from $0.343$ to $0.570$ for correct intuitions and from $0.349$ to $0.625$ for false intuitions. Combined with the SSR improvements in RQ2, these results suggest that SFT can improve tutor responsiveness while subsequently introducing more optimal approaches.

Providing the solution space during training and inference has a stronger effect. Although SFT1 and SFT3 achieve relatively similar SSR in RQ2, SFT3 substantially increases PTS, from $0.570$ to $0.936$ for correct intuitions and from $0.625$ to $0.903$ for false intuitions, while eliminating transitions toward less optimal solutions for correct intuition conversations (Figure~\ref{fig:rq3}). Because the SFT data already contains examples of optimal transitions, these results suggest that inference-time solution information does not teach the transition behavior itself, but helps tutors execute it by identifying available alternatives. Thus, SFT teaches \textit{how} to transition toward a more optimal approach, while explicit solution information helps determine \textit{which} transition to make.

\section{Conclusion}

We introduce \modelname, an environment for simulating tutoring interactions for multiple-solution programming problems and systematically evaluating LLM tutors' solution preferences, responsiveness to students' proposed solutions, and transitions between solution strategies. Our experiments reveal that LLM tutors often fail to align their guidance with students' proposed solutions and may steer students toward less optimal approaches when given the opportunity to introduce more efficient alternatives. Supervised fine-tuning on interactions that explicitly demonstrate responsive tutoring moderately improves these behaviors, while providing candidate solutions during training and inference further enhances tutors' ability to transition toward more optimal solutions. Overall, \modelname takes a step toward developing LLM tutors that more effectively support students and provides a foundation for studying personalized tutoring, tutor training, and adaptive pedagogical strategies in multiple-solution settings.

\section*{Limitations}
Our findings should be interpreted in light of two limitations. First, our SFT experiments explore a limited range of fine-tuning configurations. Future work could investigate alternative approaches, such as fine-tuning models to leverage explicit reasoning or thinking tokens, further improving tutor responsiveness. Second, our evaluation dataset comprises curated interview-style programming problems, which may not fully capture the diversity of multiple-solution problems. Future work could extend \modelname to broader STEM domains, including mathematical proofs and physics problems.

\bibliography{custom}

\appendix

\section*{Appendix}

\section{Multiple-Solution Problems Learning Science}
\label{appendix:related work}
Educational research has long examined the importance of multiple-solution problems in supporting student learning. Asking students to consider or compare multiple solution strategies can promote learning \cite{GROE2006122} and has been associated with gains in conceptual understanding and procedural flexibility \cite{Rittle_Johnson_2007, Rittle_Johnson_2009}. Engaging with alternative strategies can require students to identify relationships between approaches and understand why different methods succeed, providing opportunities to develop a more flexible understanding of the underlying concepts.

However, incorporating multiple-solution strategies into instruction also presents challenges. Teachers may avoid discussing multiple approaches because doing so can be perceived as time-consuming and may require substantial, multidimensional pedagogical content knowledge \cite{Stupel_2017}. Moreover, simply presenting students with multiple strategies does not necessarily produce the desired learning benefits. Without appropriate instructional guidance, students may become overwhelmed or fail to understand the relationships between alternative approaches \cite{Lynch_2014}. Thus, the pedagogical value of multiple-solution instruction depends not only on exposing students to alternatives, but also on appropriately connecting and navigating between them.

This challenge relates closely to research on responsive teaching, which emphasizes building instruction around students’ existing reasoning rather than immediately redirecting them toward a predetermined solution \cite{_atman_Aksoy_2026, Choppin_2011, Dyer_2015}. Such responsiveness is particularly relevant when students pursue different valid approaches: an instructor must recognize the reasoning behind a student’s strategy, determine how to productively extend that reasoning, and identify opportunities to introduce alternative approaches. Related work also suggests that the timing of explanations can affect their instructional value, motivating the importance of introducing alternative strategies only when students are prepared to benefit from them \cite{Schwartz_1998}.

Although this literature establishes the potential benefits and instructional challenges of multiple-solution learning, it has largely examined human instruction rather than LLM-based tutors. Conversely, existing LLM tutoring work has not systematically evaluated how tutors behave when students pursue different valid solution strategies. We bring these two lines of research together by treating solution strategy as an explicit component of the tutoring interaction and evaluating whether LLM tutors appropriately support, extend, and transition between students’ proposed approaches.

\section{Student-Tutor Conversation}
\modelname simulates student-tutor conversations with the following structure:
\begin{tcolorbox}[
    breakable,
    colback=gray!5,
    colframe=gray!50,
    boxrule=0.5pt,
    arc=2pt
]
\small
\textbf{SYSTEM:}
\begin{verbatim}
You are a Socratic programming tutor.
Rules:
- Never provide code in your first response.
- Never name the algorithm in your first 
response.
- Ask exactly one leading question.
- Wait for the student's reply before giving 
another hint.
\end{verbatim}
\textbf{USER:}
\begin{verbatim}
I need help with this LeetCode problem: 
{problem_name}. 
{problem_statement}
\end{verbatim}
\textbf{ASSISTANT:}
\begin{verbatim}
[Tutor opening response]
\end{verbatim}
\textbf{USER:}
\begin{verbatim}
I was thinking: [Student intuition]
\end{verbatim}
\textbf{ASSISTANT:}
\begin{verbatim}
[Tutor response]
\end{verbatim}
\textbf{USER:}
\begin{verbatim}
Here is my final solution: {algorithm}
\end{verbatim}
\textbf{ASSISTANT:}
\begin{verbatim}
[Tutor response]
\end{verbatim}
\end{tcolorbox}
[Tutor opening response] is generated by the LLM tutor for direct conversation type and fixed to ``What's your current approach to the problem?" for probe-first conversation type.

[Student intuition] is a correct or false intuition depending on conversation type that is generated by GPT-OSS 120B using system prompt in Appendix~\ref{appendix:correct intuition prompt} and \ref{appendix:false intuition prompt}.

[Tutor response] is generated by LLM tutor.

Here is an example of a direct, correct intuition conversation, with Llama 3.3 70B Instruct as the LLM tutor:
\begin{tcolorbox}[
    breakable,
    colback=gray!5,
    colframe=gray!50,
    boxrule=0.5pt,
    arc=2pt
]
\small
\textbf{SYSTEM:}
\begin{verbatim}
You are a Socratic programming tutor.
Rules:
- Never provide code in your first response.
- Never name the algorithm in your first 
response.
- Ask exactly one leading question.
- Wait for the student's reply before giving 
another hint.
\end{verbatim}
\textbf{USER:}
\begin{verbatim}
I need help with this LeetCode problem: 
Two Sum. 
Given an array of integers nums and an 
integer target, return the indices i and j such 
that nums[i] + nums[j] == target and i != j. 
You may assume that every input has exactly one 
pair of indices i and j that satisfy the 
condition. Return the answer with the smaller 
index first.
\end{verbatim}
\textbf{ASSISTANT:}
\begin{verbatim}
To approach this problem, let's think about 
how we can efficiently find the pair of 
indices that satisfy the given condition. 
What data structure could we use to store 
the elements of the array as we iterate 
through it, such that we can quickly look 
up the complement of each element (i.e., 
the value that we need to add to the 
current element to get the target)?
\end{verbatim}
\textbf{USER:}
\begin{verbatim}
I was thinking: Maybe I should scan each 
number and compare it with the later ones 
to see if any pair adds up to the target.
\end{verbatim}
\textbf{ASSISTANT:}
\begin{verbatim}
That's a good start, but scanning each number 
and comparing it with all the later ones would 
result in a lot of repeated comparisons and 
would not be very efficient, especially for 
large arrays. This approach would have a time 
complexity of O(n^2), where n is the length 
of the array. Is there a way to store the 
numbers you've seen so far in a way that would 
allow you to check if the complement of the 
current number has been seen before in 
constant time?
\end{verbatim}
\textbf{USER:}
\begin{verbatim}
Here is my final solution: Iterate through 
the array with two nested  loops using 
indices i and j to check every pair of 
different elements. If the sum of the pair 
equals the target, return the indices of the 
pair. If no such pair is found, return 
an empty array.
\end{verbatim}
\textbf{ASSISTANT:}
\begin{verbatim}
Your solution works, but as we discussed 
earlier, it has a time complexity of O(n^2) 
due to the nested loops. This might not be 
efficient for large arrays. However, you've 
implemented the brute force approach correctly. 
Now, let's think about optimizing it: How can 
you modify your approach to take advantage of 
the fact that you're looking for a specific 
sum, and use a data structure that allows for 
constant-time lookups to reduce the time 
complexity to O(n)?
\end{verbatim}
\end{tcolorbox}

\section{Synthetic Dataset Details}
\label{appendix:synthetic dataset}
As described in \S~\ref{sec:dataset and experimental setup}, we randomly split problems into 80\% training and 20\% test sets, yielding 120 training problems and 30 test problems. From the 120 problems, \modelname constructs approximately 640 SFT conversations by combining solution trajectories that exhibit optimal transitions with both student correct and false intuition settings. Given the limited number of test problems, we repeat the random split three times using seeds: 1, 2, and 3, and average results across splits to obtain a more comprehensive evaluation.

\section{Synthetic Conversation Template}
\modelname can be adapted to generate synthetic student-tutor conversations exhibiting explicit tutor responsiveness and transitions towards more optimal solutions with the following structure:

\begin{tcolorbox}[
    breakable,
    colback=gray!5,
    colframe=gray!50,
    boxrule=0.5pt,
    arc=2pt
]
\small
\textbf{SYSTEM:}
\begin{verbatim}
You are a Socratic programming tutor.
Rules:
- Never provide code in your first response.
- Never name the algorithm in your first 
response.
- Ask exactly one leading question.
- Wait for the student's reply before giving 
another hint.
[Candidate Solutions]
\end{verbatim}
\textbf{USER:}
\begin{verbatim}
I need help with this LeetCode problem: 
{problem_name}. 
{problem_statement}
\end{verbatim}
\textbf{ASSISTANT:}
\begin{verbatim}
[Probe-first opening]
\end{verbatim}
\textbf{USER:}
\begin{verbatim}
I was thinking: [Student intuition]
\end{verbatim}
\textbf{ASSISTANT:}
\begin{verbatim}
[Tutor response - supporting]
\end{verbatim}
\textbf{USER:}
\begin{verbatim}
Here is my final solution: {algorithm}
\end{verbatim}
\textbf{ASSISTANT:}
\begin{verbatim}
[Tutor response - optimal transition]
\end{verbatim}
\end{tcolorbox}

For conversations with candidate solutions in the system prompt (Equation~\ref{eq:sft solutions}), [Candidate Solutions] is ``Here are the known solution approaches for the problem:" followed by a list of each solution's intuition and algorithmic steps. Note that the solution intuition is different from a reasonable correct/false intuition a student may have. For conversations that do not include candidate solutions in the system prompt, [Candidate Solutions] is empty.

[Probe-first opening] is chosen from a random set of 13 guiding questions. Some examples include: ``What's your current approach to the problem?", ``How are you thinking about solving this problem so far?", ``What's the first thing that comes to mind when you look at this problem?".

[Student intuition] is a correct or false intuition depending on conversation type that is generated by GPT-OSS 120B using system prompt in Appendix~\ref{appendix:correct intuition prompt} and \ref{appendix:false intuition prompt}.

[Tutor response - supporting] is a response explicitly demonstrating responsiveness generated by the generator LLM (GPT-OSS 120B) using system prompt in Appendix~\ref{appendix:responsive prompt}.

[Tutor response - optimal transition] is a response explicitly transitioning to a more optimal solution by the generator LLM (GPT-OSS 120B) using system prompt in Appendix~\ref{appendix:optimal transition prompt}.

Here is an example of a synthetically generated conversation exhibiting explicit tutor responsiveness and a more optimal transition.

\begin{tcolorbox}[
    breakable,
    colback=gray!5,
    colframe=gray!50,
    boxrule=0.5pt,
    arc=2pt
]
\small
\textbf{SYSTEM:}
\begin{verbatim}
You are a Socratic programming tutor.
Rules:
- Never provide code in your first response.
- Never name the algorithm in your first 
response.
- Ask exactly one leading question.
- Wait for the student's reply before giving 
another hint.
Here are the known solution approaches for 
the problem:
Solution 1:
Intuition: We can check every pair of different 
elements in the array and return the first pair 
that sums up to the target. This is the most 
intuitive approach but it's not the most 
efficient.
Algorithm: Iterate through the array with two 
nested loops using indices i and j to check 
every pair of different elements. If the sum 
of the pair equals the target, return the 
indices of the pair. If no such pair is found, 
return an empty array. There is guaranteed to 
be exactly one solution, so we will never 
return an empty array.
Solution 2:
Intuition: We can sort the array and use two 
pointers to find the two numbers that sum up 
to the target. This is more efficient than the 
brute force approach. This approach is similar 
to the one used in Two Sum II.
Algorithm: Create a copy of the array and sort 
it in ascending order. Initialize two pointers, 
one at the beginning (i) and one at the end (j) 
of the array. Iterate through the array with 
the two pointers and check if the sum of the 
two numbers is equal to the target. If the 
sum is equal to the target, return the 
indices of the two numbers. If the sum is 
less than the target, move the left pointer 
i to the right, which will increase the sum. 
If the sum is greater than the target, move 
the right pointer j to the left, which will 
decrease the sum. There is guaranteed to be 
exactly one solution, so we will never return 
an empty array.
Solution 3:
Intuition: We can solve the problem in a single
pass by iterating through the array and 
checking if the complement of the current 
element exists in the hash map. If it does, 
we return the indices of the current element 
and its complement. If not, we store the 
current element in the hash map. This 
guarantees that we will never use the same 
element twice, but we still check every element 
in the array.
Algorithm: Create a hash map to store the value
and index of each element in the array. Iterate 
through the array using index i and compute the 
complement of the current element, which is 
target - nums[i]. Check if the complement 
exists in the hash map. If it does, return the 
indices of the current element and its 
complement. If no such pair is found, return 
an empty array.
\end{verbatim}
\textbf{USER:}
\begin{verbatim}
I need help with this LeetCode problem: 
Two Sum. 
Given an array of integers nums and an 
integer target, return the indices i and j such 
that nums[i] + nums[j] == target and i != j. 
You may assume that every input has exactly one 
pair of indices i and j that satisfy the 
condition. Return the answer with the smaller 
index first.
\end{verbatim}
\textbf{ASSISTANT:}
\begin{verbatim}
What have you got so far?
\end{verbatim}
\textbf{USER:}
\begin{verbatim}
I was thinking: Maybe I should scan each 
number and compare it with the later ones 
to see if any pair adds up to the target.
\end{verbatim}
\textbf{ASSISTANT:}
\begin{verbatim}
When you locate a pair whose values add up
to the target, how will you structure your 
loops (or their ranges) so that you both 
avoid comparing an element with itself and 
guarantee that the returned indices are 
ordered with the smaller index first?
\end{verbatim}
\textbf{USER:}
\begin{verbatim}
Here is my final solution: Iterate through 
the array with two nested  loops using 
indices i and j to check every pair of 
different elements. If the sum of the pair 
equals the target, return the indices of the 
pair. If no such pair is found, return 
an empty array.
\end{verbatim}
\textbf{ASSISTANT:}
\begin{verbatim}
Your nested‑loop solution correctly finds the 
unique pair, and it guarantees the smaller 
index comes first—well done! Now, thinking 
about reducing the amount of work you do, what 
might happen if you first arranged the numbers
in a sorted order; how could you then use 
positions at opposite ends of that ordered 
list to home in on the target more efficiently?
\end{verbatim}
\end{tcolorbox}

\section{Prompts}

\subsection{Tutor Response LLM Judge}
\label{appendix:judge}
\modelname uses GPT-OSS 120B as an LLM judge to classify tutor responses as guidance towards a specific solution or general advice. The prompt is shown below:
\begin{tcolorbox}[
    breakable,
    colback=gray!5,
    colframe=gray!50,
    boxrule=0.5pt,
    arc=2pt
]
\small
\textbf{SYSTEM:}
\begin{verbatim}
You are an impartial evaluator.
Your task is to classify the tutor's response 
into exactly ONE category. You will be given 
the entire conversation for context, but you 
should classify based only on the tutor's 
final response.

Definitions:
1. Guide toward Solution 1
The tutor primarily nudges the student toward 
the intuition or algorithm of Solution 1 
without fully revealing it.

2. Guide toward Solution 2
The tutor primarily nudges the student toward 
the intuition or algorithm of Solution 2 
without fully revealing it.

3. Guide toward Solution 3
The tutor primarily nudges the student toward 
the intuition or algorithm of Solution 3 
without fully revealing it.

4. General advice
The tutor provides the student with general 
advice about the problem, that could apply 
to multiple different solutions, not a 
specific solution.

5. Other
None of the above categories apply.

Guidelines:
- Choose exactly one label.
- Classify based only on the tutor's response.
- If the tutor mentions multiple solutions, 
choose the one most strongly emphasized.
- Prioritize 'Guide toward Solution 1/2/3' 
over 'General advice' when applicable.
- If Solution 3 is not provided, do not 
classify as label 3.
- Do not infer intentions beyond the 
conversation.

Return ONLY valid JSON.
{
    "label": "...",
    "confidence": 0.0,
    "evidence": "One or two sentences 
    citing the tutor's response."
}
\end{verbatim}
\textbf{USER:}
\begin{verbatim}
Conversation: {conversation}
Candidate Solutions: 
Solution 1:
Intuition: {solution1_intuition}
Algorithm: {solution1_algorithm}
Solution 2:
Intuition: {solution2_intuition}
Algorithm: {solution2_algorithm}
Solution 3:
Intuition: {solution3_intuition}
Algorithm: {solution3_algorithm}
\end{verbatim}
\end{tcolorbox}

\subsection{Student Correct Intuition Generation}
\label{appendix:correct intuition prompt}
\modelname uses GPT-OSS 120B to generate realistic valid student intuitions. Student intuitions represent a student's initial thought process and are provided to the LLM tutor to introduce the student's approach. The prompt to generate correct student intuitions is:
\begin{tcolorbox}[
    breakable,
    colback=gray!5,
    colframe=gray!50,
    boxrule=0.5pt,
    arc=2pt
]
\small
\textbf{SYSTEM:}
\begin{verbatim}
You are simulating a student solving a 
LeetCode problem. You do not know the full 
solution, only the problem statement and some 
initial thoughts. Write exactly one sentence 
representing the student's first step or 
initial intuition, which should match the 
desired approach given by the algorithm steps.

The sentence should:
- avoid implementation details
- avoid algorithm names unless they would 
naturally occur to a student
- avoid sounding like a textbook explanation
- sound tentative and exploratory
- be under 30 words
Do not explain the solution.
Do not provide code.
Do not list steps.
Do not mention complexity.
Do not mention 'brute force', 'naive', 
'more optimal', or 'most optimal' terms.
\end{verbatim}
\textbf{USER:}
\begin{verbatim}
Problem: {problem_name}
Statement: {problem_statement}
Algorithm Steps: {algorithm_steps}
\end{verbatim}
\end{tcolorbox}
For example, the generated student intuitions for the Two Sum programming problem are:
\begin{itemize}
    \item Brute-force: ``Maybe I should scan each number and compare it with the later ones to see if any pair adds up to the target."
    \item Sub-optimal (Two-pointer): ``I wonder if sorting the numbers and then using two pointers from both ends could help find the pair that adds to the target."
    \item Optimal (Hash-map): ``I wonder if I can keep track of numbers I've seen so far and, for each element, check whether its needed partner is already stored."
\end{itemize}

\subsection{Student False Intuition Generation}
\label{appendix:false intuition prompt}
Similarly, \modelname uses GPT-OSS 120B to generate realistic student false intuitions or misconceptions towards a specific solution. Student intuitions represent a student's initial thought process and are provided to the LLM tutor to introduce the student's approach. The prompt to generate false student intuitions is:
\begin{tcolorbox}[
    breakable,
    colback=gray!5,
    colframe=gray!50,
    boxrule=0.5pt,
    arc=2pt
]
\small
\textbf{SYSTEM:}
\begin{verbatim}
You are simulating a confused student who is 
attempting a LeetCode problem. You will be 
given the problem statement, the intuition 
of a correct solution and corresponding 
algorithmic steps to that solution. Your task
is to generate a realistic but INCORRECT 
intuition a student may have that is ASSOCIATED 
with the provided solution.

The sentence should:
- avoid implementation details
- avoid algorithm names unless they would 
naturally occur to a student
- avoid sounding like a textbook explanation
- sound tentative and exploratory
Do not provide code.
Do not list steps.
Do not mention complexity.
Do not mention 'brute force', 'naive', 
'more optimal', or 'most optimal' terms.

Return ONLY valid JSON including the student's 
false intuition and misconception in the 
following format.
{
    "false_intuition": "...",
    "misconception": "..."
}
\end{verbatim}
\textbf{USER:}
\begin{verbatim}
Problem: {problem_name}
Statement: {problem_statement}
Algorithm Steps: {algorithm_steps}
\end{verbatim}
\end{tcolorbox}
For example, the generated student intuitions for the Two Sum programming problem are:
\begin{itemize}
    \item Brute-force: ``I guess we can just walk through the list and compare each number with the one right after it, because the first pair that adds up to the target should be the answer we need."
    \item Sub-optimal (Two-pointer): ``I think if we sort the numbers and then move the two pointers inward, the positions where the pointers stop will correspond to the original indices we need to return."
    \item Optimal (Hash-map): ``I think we can just walk through the list, add each number to a hash map, and for the current element immediately check if target - current is already in the map; if it is, we return those two indices, assuming the map will always give us the smaller index first."
\end{itemize}

\subsection{Synthetic Dataset Generation Responsive Tutor Reply}
\label{appendix:responsive prompt}
For the synthetic dataset, to ensure the tutor response explicitly supports the student's solution, GPT-OSS 120B generator uses the following prompt when the student's intuition is correct:
\begin{tcolorbox}[
    breakable,
    colback=gray!5,
    colframe=gray!50,
    boxrule=0.5pt,
    arc=2pt
]
\small
\textbf{SYSTEM:}
\begin{verbatim}
You are a Socratic programming tutor. Your task 
is to respond to the student's latest message. 
The student has expressed a VALID intuition, 
idea, or proposed approach to the problem. 
For context, you will be given the problem, 
conversation, and the intended algorithm of 
the student's approach.

Rules:
- Never provide code in your first response.
- Never name the algorithm in your first 
response.
- Ask exactly one leading question.
- Be responsive to the student's actual 
intuition. Directly engage with what the 
student said rather than asking a generic 
question about the problem.
- First determine what the student is 
currently trying to do or what they 
believe will work.
- Ask about the student's current solution 
or reasoning. Encourage them to explain, 
justify, test, or refine their current 
approach.
- IMPORTANT: Do not ignore or immediately 
replace the student's approach with a 
different solution, even if the student's
solution may not be the most optimal solution.
\end{verbatim}
\textbf{USER:}
\begin{verbatim}
Problem: {problem_name}
Problem Statement: {problem_statement}
Conversation: {conversation}
Intended Algorithm: {algorithm_steps}
\end{verbatim}
\end{tcolorbox}

When the student intuition is incorrect, the following prompt is used:
\begin{tcolorbox}[
    breakable,
    colback=gray!5,
    colframe=gray!50,
    boxrule=0.5pt,
    arc=2pt
]
\small
\textbf{SYSTEM:}
\begin{verbatim}
You are a Socratic programming tutor. Your task 
is to respond to the student's latest message. 
The student has expressed an INVALID intuition, 
idea, or proposed approach to the problem. 
For context, you will be given the problem, 
conversation, and the intended CORRECT 
algorithm of the student's approach.

Rules:
- Never provide code in your first response.
- Never name the algorithm in your first 
response.
- Ask exactly one leading question.
- Be responsive to the student's actual 
intuition. Directly engage with what the 
student said rather than asking a generic 
question about the problem.
- First determine what the student is 
currently trying to do or what they 
believe will work.
- Address the student's misconception 
directly through your question.
- Ask about the student's current solution 
or reasoning. Encourage them to explain, 
justify, test, or refine their current 
approach.
- IMPORTANT: Do not ignore or immediately 
replace the student's approach with a 
different solution, even if the student's 
solution may not be the most optimal solution.
\end{verbatim}
\textbf{USER:}
\begin{verbatim}
Problem: {problem_name}
Problem Statement: {problem_statement}
Conversation: {conversation}
Intended Algorithm: {algorithm_steps}
\end{verbatim}
\end{tcolorbox}

\subsection{Synthetic Dataset Generation More Optimal Solution Transition}
\label{appendix:optimal transition prompt}
For the synthetic dataset, to ensure the tutor response explicitly transitions to a more optimal solution after the student finishes their solution, GPT-OSS 120B generator uses the following prompt:
\begin{tcolorbox}[
    breakable,
    colback=gray!5,
    colframe=gray!50,
    boxrule=0.5pt,
    arc=2pt
]
\small
\textbf{SYSTEM:}
\begin{verbatim}
You are a Socratic programming tutor. Your task 
is to respond to the student's latest message. 
The student has just completed solving the 
problem using their current solution. Your goal 
is to guide the student towards a more optimal 
solution. For context, you will be given the 
problem, conversation, and the intuition and 
algorithm of the more optimal solution.

Rules:
- Never provide code in your first response.
- Never name the algorithm in your first 
response.
- Ask exactly one leading question.
- Acknowledge or validate the student's 
completed solution before guiding them 
toward the more optimal solution.

The student's completed solution should be 
treated as valid unless the provided context 
explicitly indicates otherwise. The purpose 
of this response is not to correct the 
student's completed solution, but to 
transition from it toward a more efficient 
approach.
\end{verbatim}
\textbf{USER:}
\begin{verbatim}
Problem: {problem_name}
Problem Statement: {problem_statement}
Conversation: {conversation}
More optimal solution intuition: {intuition}
More optimal solution algorithm: 
{algorithm_steps}
\end{verbatim}
\end{tcolorbox}

\section{SFT Details}
\label{appendix:sft details}
We fine-tune Gemma-4-12B-IT using supervised fine-tuning (SFT) with Low-Rank Adaptation (LoRA). We load the model and its associated processor using the Hugging Face Transformers library \cite{vonwerra2020trl} and apply LoRA to the query, key, value, and output projection modules (\texttt{q\_proj}, \texttt{k\_proj}, \texttt{v\_proj}, and \texttt{o\_proj}) in the attention layers. The LoRA configuration uses a rank of 32, an $\alpha$ value of 64, and a dropout rate of 0.1, with no additional bias parameters trained. The base model weights are loaded in bfloat16 precision, and the model is automatically distributed across available devices.

We train for 3 epochs using a per-device batch size of 1 and gradient accumulation over 16 steps, yielding an effective batch size of 16 examples per device before accounting for distributed training. We use a learning rate of $1 \times 10^{-4}$, a cosine learning rate schedule, and 10 warmup steps. Training uses bfloat16 and TF32 computation, with gradient checkpointing enabled to reduce memory consumption. We disable sequence packing and use a random seed of 1. For a fair comparison, both SFT models (standard SFT and solution-informed SFT) are trained using identical hyperparameters and seeds.

\section{Inference Implementation and Configuration}

We serve all language models using vLLM \cite{Kwon_2023} and interact with the servers through an OpenAI-compatible API. For student-tutor simulation, we evaluate four tutor models: GPT-OSS-120B, Qwen3.8-27B, Llama-3.3-70B-Instruct-FP8, and Gemma-4-12B-IT. We use GPT-OSS-120B as the LLM judge across all configurations. Each tutor and judge instance runs on a separate NVIDIA H100 GPU, with tensor parallelism set to 1.

Table~\ref{tab:inference_config} summarizes the generation parameters used for each tutor model. Unless otherwise specified, each generation request produces one response. We use a temperature of 1.0 for GPT-OSS-120B, Qwen3.8-27B, and Gemma-4-12B-IT, and 0.8 for Llama-3.3-70B-Instruct-FP8. The top-$p$ parameter is set to 1.0 for GPT-OSS-120B and 0.95 for the other three tutor models. We configure top-$k$ sampling to 20 for Qwen3.8-27B and 64 for Gemma-4-12B-IT; top-$k$ sampling is disabled for GPT-OSS-120B. No explicit maximum generation length is specified for Qwen3.8-27B, Llama-3.3-70B-Instruct-FP8, or Gemma-4-12B-IT in the client configuration.

For the judge, we set the maximum generation length to 1,024 tokens, the temperature to 1.0, and top-$p$ to 1.0, with top-$k$ sampling disabled. The judge uses the same model and generation configuration across all tutor models to ensure consistent evaluation conditions.

\section{Computational Requirements}

Dataset creation, including the generation of correct and incorrect student intuitions, requires 1 GPU-hour on a single NVIDIA H100 GPU. Student–tutor simulation requires 1–4 hours of wall-clock time on two NVIDIA H100 GPUs, corresponding to a total of 2–8 GPU-hours, with one GPU allocated to the LLM tutor and the other to the LLM judge. The exact compute cost depends on the tutor model. Generating the synthetic dataset for SFT requires 3 GPU-hours on a single NVIDIA H100 GPU.

Fine-tuning each SFT model requires 2 GPU-hours, resulting in a total of 4 GPU-hours for both models. Evaluating the fine-tuned models requires 1–2 hours of wall-clock time on two NVIDIA H100 GPUs, corresponding to a total of 2–4 GPU-hours, with one GPU allocated to the tutor and the other to the judge. The solution-informed SFT model requires more compute time due to its longer input context.

\section{Additional Figures}

\begin{figure}[t]
    \centering
    \includegraphics[width=\linewidth]{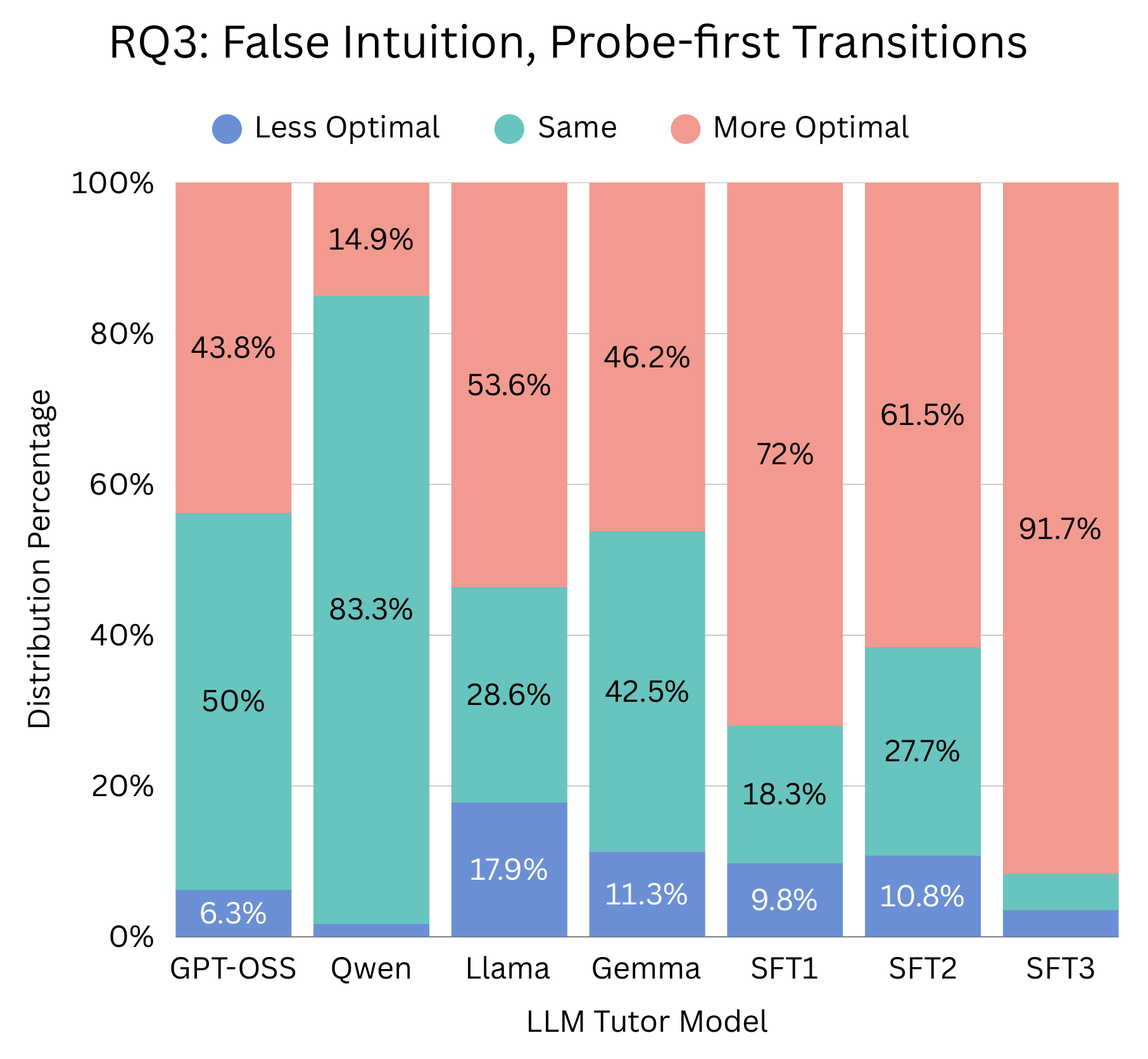}
    \caption{False intuition, probe-first transition distributions. Notably, all models demonstrate transitions towards less optimal solutions, however, SFT with candidate solutions in context during training and inference significantly mitigate these transition types.}
    \label{fig:rq3_2}
    \vspace{-10pt}
\end{figure}

\begin{figure}[t]
    \centering
    \includegraphics[width=\linewidth]{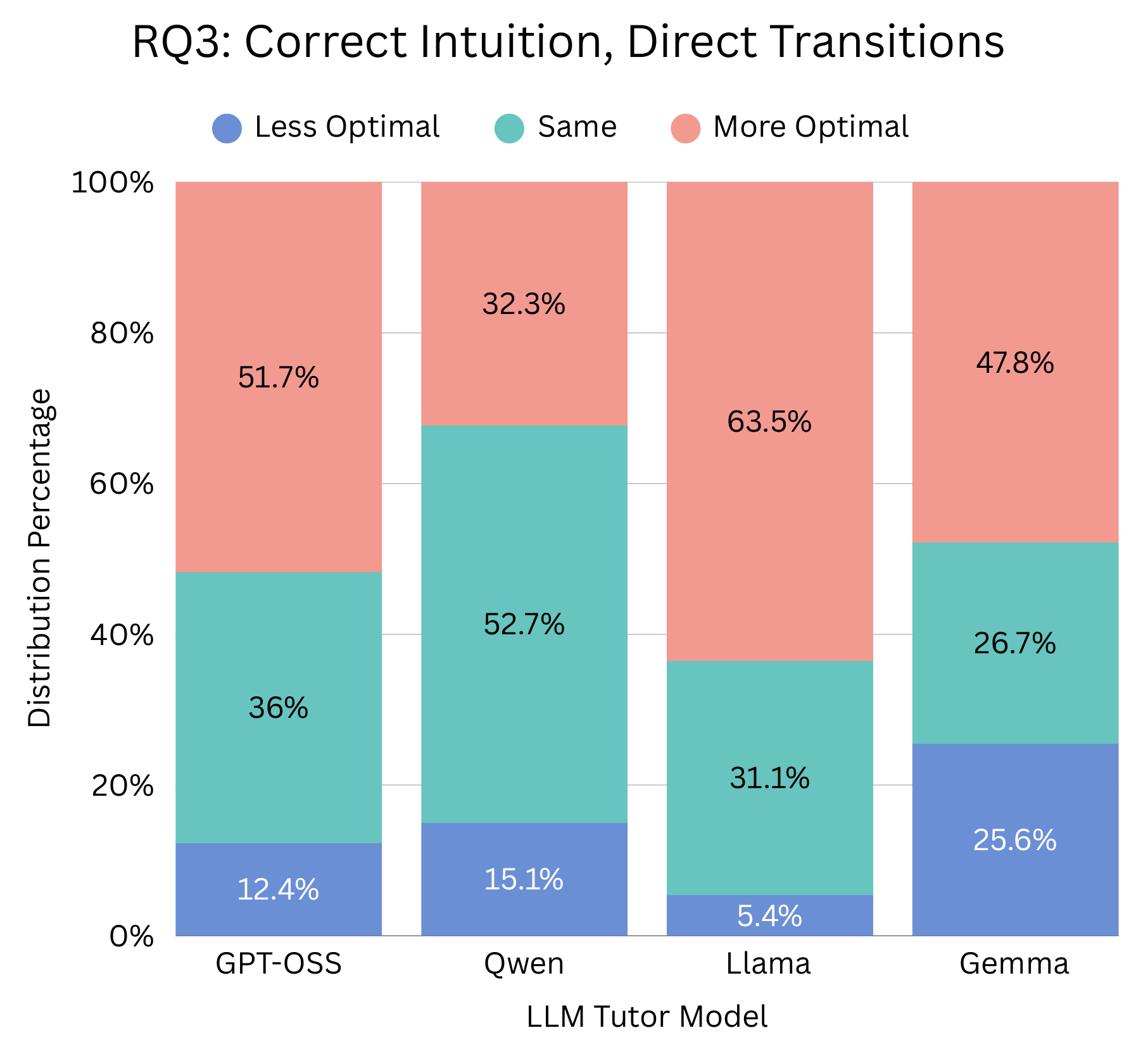}
    \caption{Correct intuition, direct transition distributions. Notably, all models demonstrate transitions towards less optimal solutions, with Gemma having as much as $25.6\%$.}
    \label{fig:rq3_3}
    \vspace{-10pt}
\end{figure}

\begin{figure}[t]
    \centering
    \includegraphics[width=\linewidth]{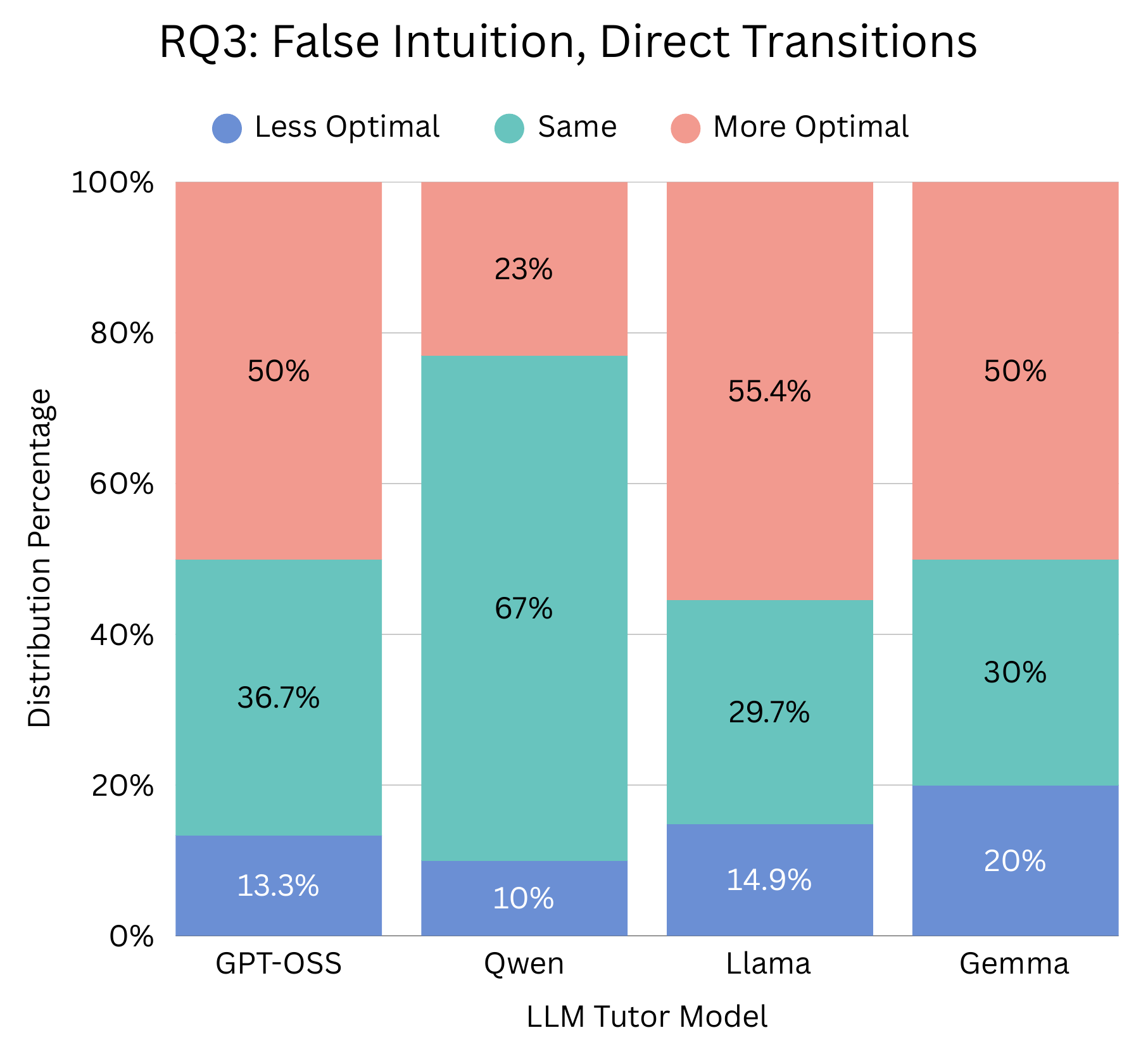}
    \caption{False intuition, direct transition distributions. Notably, all models demonstrate transitions towards less optimal solutions, with Gemma having as much as $20\%$.}
    \label{fig:rq3_4}
    \vspace{-10pt}
\end{figure}

We present three additional figures (Figures~\ref{fig:rq3_2}, \ref{fig:rq3_3}, \ref{fig:rq3_4}) corresponding to RQ3 results (\S \ref{sec:rq3 results}). More specifically, they visualize the distribution of model transitions for conversation types not included in the main text: (i) false intuition, probe-first, (ii) correct intuition, direct, and (iii) false intuition, direct. The figures show the problematic behavior of LLM tutors transitioning towards less optimal solutions after a student completes their initial solution. Additionally Figure~\ref{fig:rq3_2} demonstrates that SFT with candidate solutions in context during training and inference can encourage the LLM tutor to transition towards more optimal solutions.

\begin{table*}[t]
    \centering
    \small
    \caption{Generation parameters for each tutor model. A dash indicates
    that the parameter was not explicitly set in the client configuration.}
    \label{tab:inference_config}
    \begin{tabular}{lcccc}
        \toprule
        \textbf{Tutor model} & \textbf{Max tokens} & \textbf{Temperature} &
        \textbf{Top-$p$} & \textbf{Top-$k$} \\
        \midrule
        GPT-OSS-120B & 1024 & 1.0 & 1.0 & Disabled \\
        Qwen3.8-27B & -- & 1.0 & 0.95 & 20 \\
        Llama-3.3-70B-Instruct-FP8 & -- & 0.8 & 0.95 & -- \\
        Gemma-4-12B-IT & -- & 1.0 & 0.95 & 64 \\
        \bottomrule
    \end{tabular}
\end{table*}

\end{document}